\documentclass[a4paper]{article}

\usepackage[english]{babel}
\usepackage[T1]{fontenc}
\usepackage{multirow}

\usepackage[a4paper,top=3cm,bottom=2cm,left=3cm,right=3cm,marginparwidth=1.75cm]{geometry}

\usepackage{amsmath}
\usepackage{graphicx}
\usepackage[colorinlistoftodos]{todonotes}
\usepackage[colorlinks=true, allcolors=blue]{hyperref}

\title{Digital holographic particle volume reconstruction \\
using a deep neural network}
\author{Tomoyoshi Shimobaba\footnote{Graduate School of Engineering, Chiba University, 1-33, Yayoi-cho, Inage-ku, Chiba, 263-8522, Japan},
Takayuki Takahashi\footnotemark[1], 
Yota Yamamoto\footnotemark[1], \\
Yutaka Endo\footnote{Institute of Science and Engineering, Kanazawa University,  Kakuma, Kanazawa, Japan, 920-1192}, 
Atsushi Shiraki\footnote{Institute of Management and Information Technologies, Chiba University, 1-33, Yayoi-cho, Inage-ku, Chiba, 263-8522, Japan}, 
Takashi Nishitsuji\footnote{Faculty of Systems Design, Tokyo Metropolitan University, 6-6, Asahigaoka, Hino-shi, Tokyo, 191-0065, Japan}, \\
Naoto Hoshikawa\footnote{Department of Innovative Electrical and Electronic Engineering, National Institute of Technology, Oyama College, 771 Nakakuki, Oyama-shi, Tochigi, 323-0806, Japan }, 
Takashi Kakue\footnotemark[1], and 
Tomoyosh Ito\footnotemark[1]}

\begin{document}
\maketitle

\begin{abstract}
This paper proposes a particle volume reconstruction directly from an in-line hologram using a deep neural network. Digital holographic volume reconstruction conventionally uses multiple diffraction calculations to obtain sectional reconstructed images from an in-line hologram, followed by detection of the lateral and axial positions, and the sizes of particles by using focus metrics. However, the axial resolution is limited by the numerical aperture of the optical system, and the processes are time-consuming.
The method proposed here can simultaneously detect the lateral and axial positions, and the particle sizes via a deep neural network (DNN).
We numerically investigated the performance of the DNN in terms of the errors in the detected positions and sizes.
The calculation time is faster than conventional diffracted-based approaches.
\end{abstract}

\section{Introduction}
Measurements of particles are used in particle image velocimetry, droplet and bubble measurements, and in particle pollution measurements in the natural environment.
Particle measurements are also used in the analysis and design of fluid flow channels.
Although the two-dimensional measurement of particles is useful, three-dimensional (3D) measurements are more valuable.
Among 3D measurement techniques, digital holographic measurements attract attention because digital holography can perform volume reconstruction of a particle field using diffraction calculations from only a single hologram \cite{adams1997particle}.

In digital holographic approaches, template matching methods have been proposed, in which the lateral and axial positions, and the size of multiple particles, are measured by template matching a measured hologram with simulated holograms \cite{murata2000potential,soulez2007inverse}.
Digital holography is capable of obtaining the phase information of an object wave. Phase information improves the precision of measuring the axial position of particles \cite{pan2003digital,ohman2016off}. The volumetric reconstruction of particles using deconvolution and a phase retrieval algorithm has been proposed \cite{latychevskaia2014holographic,tanaka2016phase}.

Recently, machine learning techniques including support vector machine (SVM), neural network and deep learning have been used in a wide range of research fields. 
Machine learning-based particle measurements have also been proposed \cite{yevick2014machine,schneider2016fast}. 
The size, refractive index and position of a single particle have been measured using SVM \cite{yevick2014machine}. First, the SVM was trained using simulated particle holograms based on Lorenz-Mie scattering. After the training,  the SVM could predict the characteristics of a single particle in a micro-channel.
The characteristics of the core and shell diameters of a single particle were determined by the shallow neural network \cite{schneider2016fast}.

In this study, we propose particle volume reconstruction directly from an in-line hologram using a deep neural network (DNN).
In general, digital holographic volume reconstruction requires multiple diffraction calculations to obtain sectional reconstructed images from an in-line hologram, followed by detection of the lateral and axial positions of the particles by using focus metrics.
These processes are time-consuming.
In addition, an in-line hologram is often contaminated by unwanted images that can interfere with the prediction.
The method proposed here is not affected by unwanted images and can simultaneously detect the lateral and axial positions, and particle sizes via the DNN.
We numerically investigated the performance of the DNN in terms of the errors of the detected positions and sizes.
The calculation time is faster than conventional diffracted-based approaches.

Section 2 describes the proposed volume reconstruction using the DNN, and Section 3 examines the performance of the proposed method. The final section concludes this work.

\section{Proposed method}

\begin{figure}
\centering
\includegraphics[width=8cm]{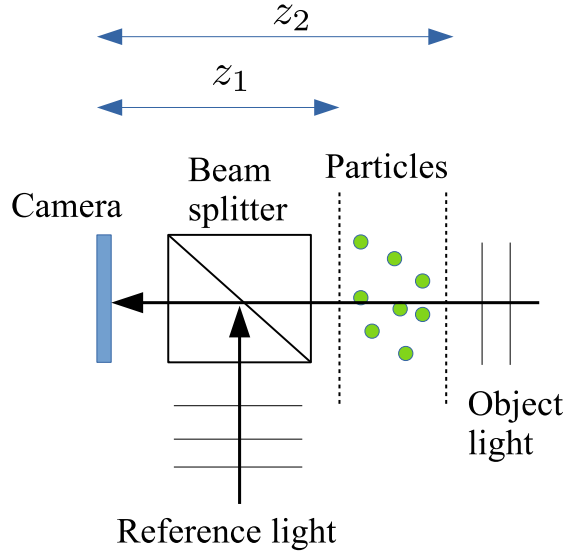}
\caption{Optical setup for recording an in-line hologram.}
\label{fig:system}
\end{figure}

In this study, we simulated particle holograms to investigate the effectiveness of a DNN that predicts the position and size of particles directly from a hologram.
We used the optical setup of an in-line hologram as shown in Fig. \ref{fig:system}.
The advantages of the in-line hologram are that the optical setup is simpler than an off-axis hologram and the space-bandwidth product of a camera can be fully utilized.
However, it is more difficult to predict particles using the in-line hologram than the off-axis hologram since the reconstructed image from the in-line hologram suffers from unwanted images (direct and conjugate lights) \cite{ohman2016off}.
The proposed method is not affected by the unwanted images and so can predict particle information. 

Holograms are calculated by the light from particles.
Light diffracted from $j$-th particle \cite{soulez2007inverse} with a position $(x_j,y_j,z_j)$ and a radius $r_j$ on a hologram plane $(x,y)$, is expressed as
\begin{equation}
u_j(x,y)=\frac{r_j}{2 i \rho_j(x-x_j,y-y_j)}{\rm J_1} \left( \frac{2 \pi r_j \rho_j(x-x_j,y-y_j)}{\lambda z_j} \right) \exp \left( \frac{i \rho^2_j(x-x_j,y-y_j)}{ \lambda z_j} \right),
\end{equation}
where $i=\sqrt{-1}$, $\rho_j(x-x_j,y-y_j)=\sqrt{(x-x_j)^2+(y-y_j)^2}$, ${\rm J_1}$ is the first-order Bessel function. 
An in-line hologram $I(x,y)$ formed by $P$ particles is calculated by
\begin{equation}
I(x,y)=\left|R(x,y)+\sum^P_{j=1} u_j(x,y)\right|^2,
\end{equation}
where $R$ is the in-line reference light.

To train the DNN, we prepared training datasets composed of holograms and particle information.
The particle information is expressed by three two-dimensional maps that indicate lateral positions, axial positions, and the sizes of the particles.
The lateral position map is expressed by a binary value; the white pixels denote the lateral positions of the particles and the black pixels denote  the absence of particles.   
The axial and size maps are expressed as 256-grayscale images. 
For example, when the axial range of particles is $z_1$ to $z_2$ (as shown in Fig. \ref{fig:system}), and the size range is $s_1$ to $s_2$, the axial and size steps per gradation are $|z_2-z_1|$ / 256 and $|s_2 - s_1|$ / 256, respectively.    

\begin{figure}
\centering
\includegraphics[width=14cm]{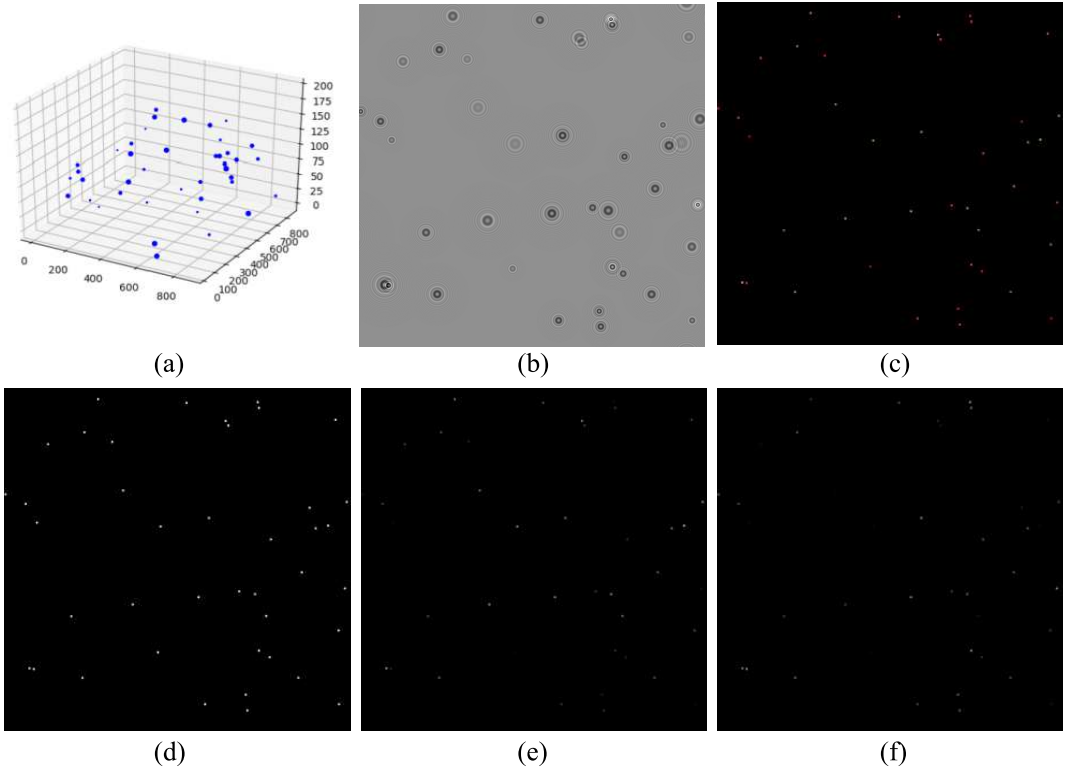}
\caption{Exemplary figures in a dataset; (a) original particle volume, (b) in-line hologram calculated from (a), (c) particle information represented by a color image, and (d)-(f) indicate the lateral and axial position maps, and size map, respectively.}
\label{fig:particle_holo}
\end{figure}

Exemplary figures of the dataset are shown in Fig. \ref{fig:particle_holo}.
Figure \ref{fig:particle_holo}(a) shows the original particle distribution in a volume, Fig. \ref{fig:particle_holo}(b) shows the in-line hologram calculated from Fig. \ref{fig:particle_holo}(a), Fig. \ref{fig:particle_holo}(c) shows the particle information represented by a color image in which the red, green and blue channels represent the lateral positions of Fig. \ref{fig:particle_holo}(d) , axial positions of Fig. \ref{fig:particle_holo}(e) and the sizes of Fig. \ref{fig:particle_holo}(f).
The particles are expressed as small squares of $5 \times 5$ pixels.
Ideally, the particles are represented by a single pixel. However, after training, the DNN did not converge well when we used the single pixel representation. 

\subsection{Deep neural network for predicting particle information}
We used U-net as the network structure for the DNN, which is used for segmentation problems \cite{ronneberger2015u} and image generation problems \cite{goodfellow2014generative}.
The prediction of the particle information can be considered as an image generation problem because the input hologram is mapped to the corresponding two-dimensional images such as those shown in Figs. \ref{fig:particle_holo}(d)-(f).
The reason why we selected the U-Net is that in a preliminary experiment, the U-Net performed better than other DNNs  (Convolutional deep auto-encoder and ResNet). 
In holography, ResNet showed  performed well at solving, problems e.g. phase unwrapping problem which can also be considered as an image generation problem mapping input holograms onto unwrapped phase maps \cite{dardikman2018phase}, but in the prediction of particle information, the training of this network was not well converged.

\begin{figure}
\centering
\includegraphics[width=\linewidth]{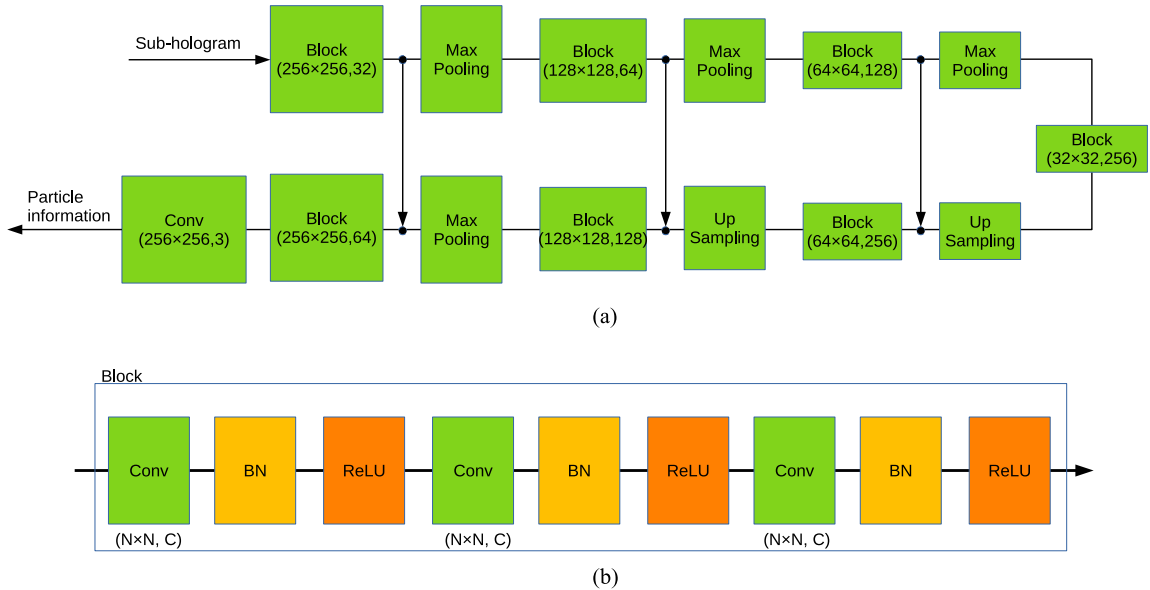}
\caption{Network structure of the DNN.}
\label{fig:dnn}
\end{figure}

The DNN  used in this study is shown in Fig. \ref{fig:dnn}.
This DNN is composed of cascading ``Block'', ''Max Pooling'' and ''Up Sampling'' layers.
The block layer has sub-layers as shown in Fig. \ref{fig:dnn}(b).
The notation $(N \times N, ~C)$ denotes the image size of $N \times N$ pixels and the number of filers $C$ for convolution layers.
``Conv'' stands for a convolution layer in which all the filter sizes are $3 \times 3$ except for the last convolution layer (the filter size of the last convolution layer is  $1 \times 1$), ''BN'' stands for the batch normalization that helps to converge the training and avoids the over-fitting problem, and ''ReLU'' stands for the activation layer using the rectified linear function. 

In Fig. \ref{fig:dnn}(a), the last layer is the convolution layer with $(256 \times 256, ~3)$ because the last layer outputs the predictions of the lateral and axial position maps, and size map.
To avoid the decrease in the detail features acquired by the upper blocks via the max pooling processes, the skip connections between the upper and bottom blocks transmit the detail features acquired by the upper blocks to the bottom ones.
The detail of the particle information is retained.

\begin{figure}
\centering
\includegraphics[width=7cm]{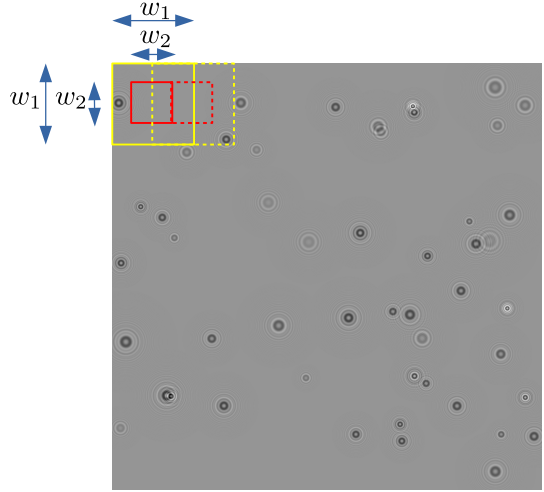}
\caption{Hologram and sub-holograms.}
\label{fig:crop}
\end{figure}

The size of a hologram is 1,024 $\times$ 1,024.
The size is too large to input the hologram directly to the DNN.
Therefore, as shown in Fig. \ref{fig:crop}, we need to divide the hologram into $w_1 \times w_1$ sub-holograms, and the sub-holograms are input to the DNN.
The size of the predicted result from the DNN is the same as the sub-hologram with $w_1 \times w_1$ pixels.
After the prediction, we extract a small region with $w_2 \times w_2$ pixels from the predicted result.
The border of the sub-holograms include incomplete interference patterns, which means the prediction in the border includes unreliable results. Therefore, the extraction of a small region gives a better-predicted result.

The prediction of the sub-hologram (the dashed boxes in Fig. \ref{fig:crop}) was then obtained by sliding the window with $w_2$ pixels. 
By repeating this process, the entire prediction result can be obtained by splicing the sub-prediction results.
We used $w_1 \times w_1 = 256 \times 256$ pixels and $w_2 \times w_2 = 128 \times 128$ pixels. Therefore, the final prediction has $896 \times 896$ pixels. 

We used the following loss function for the training:
\begin{equation}
||Y-X||^2_2 + \alpha ||X||_1,
\end{equation}
where $Y$ and $X$ indicate the original and predicted particle information, $|| \cdot ||_2$ and $|| \cdot ||_1$ denote the $\ell_2$ and $\ell_1$ norms, respectively.
The $\ell_2$ norm decreases the error between the original particle information and the predicted one, and the $\ell_1$ norm mitigates miss prediction in the black area (the values are zero) in Figs. \ref{fig:particle_holo}(c)-(f), where $\alpha$ was a hyperparameter. In this study, the hyperparameter was empirically decided as $\alpha=0.01$.

In the training process of the DNN, we needed to prepare a large number of datasets. 
The network parameters in the DNN were optimized by minimizing the loss function.
We used Adam \cite{Kingma2014Adam}, which is a stochastic gradient descent (SGD) method to minimize the loss function. 
In the SGD, we randomly selected $B$ datasets from among all of the datasets.
$B$ is referred to as the batch size, which in the case was 20.

\subsection{Post processing}
\begin{figure}
\centering
\includegraphics[width=12cm]{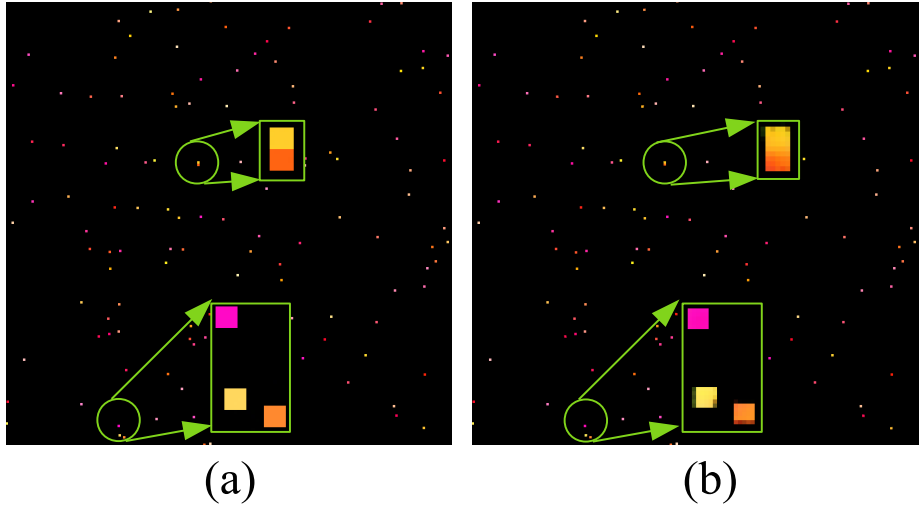}
\caption{Exemplary figures of original and predicted particle information: (a) original and (b) predicted particle information.}
\label{fig:predic_map}
\end{figure}

In the post-processing, we visualize the volume reconstruction from the predicted particle information, including the lateral and axial position maps, and the size map of the particles.
Figures \ref{fig:predic_map}(a) and \ref{fig:predic_map}(b) show exemplary images of the original and predicted particle information, respectively.
As shown in Fig.\ref{fig:predic_map}(a), the size of the squares is just 5$\times$5 pixels and the pixel values in the square are uniform. 
However, as shown in Fig. \ref{fig:predic_map}(b), although some squares are correctly predicted, some squares change slightly more than $5 \times 5$ pixels and pixel values are not uniform, due to prediction error.

To reconstruct the particle volume from the predicted particle information, first, we detect the lateral position of the particles. 
For detection, we searched for a pixel in the lateral position map with over the value of $255 \times 0.95$ by raster scan. The value 0.95 means the probability whether the pixel is the particle. 
If the pixel is found, we count the non-zero pixels in a region with $5 \times 5$ pixels starting from the position of a found pixel. 
At the same time, we calculated the median values within the region in the axial and size maps. We dealt with these median values as the axial position and size of the particle. 
When the count value exceeded $5\times5/2$, we regarded the center of the region as the lateral position of the particle.

\section{Results}
We evaluated the performance of the DNN in terms of its precision and the calculation time.
We used a computer with an Intel Xeon E5 CPU, memory of 128 GB, and a graphics processing unit (NVIDIA GeForce GTX980).
We prepared the training datasets including 9,000 holograms and their corresponding particle information, and the validation datasets with 1,000 holograms and their corresponding particle information. 
In these datasets, the number of particles was randomly changed from 50 to 200.
The number of epoch for the training was 40 and the training time was approximately two and a half days.

As shown in Fig.\ref{fig:system}, the axial positions of each particle were randomly changed within $z_1=1$ cm to $z_2=3$ cm.
The holograms have 1,024$\times$1,024 pixels and a pixel pitch of 10 $\mu$m. Therefore, the range of lateral positions of the particles is approximately 10 mm$\times$10 mm.
The sizes of particles are also randomly changed
20 $\mu$m to 100 $\mu$m.
The wavelength of the reference light was 633 nm.
These datasets were generated from our wave optics library \cite{shimobaba2012computational}.
We used Keras as the deep-learning framework \cite{chollet2015keras}.

\begin{figure}
\centering
\includegraphics[width=\linewidth]{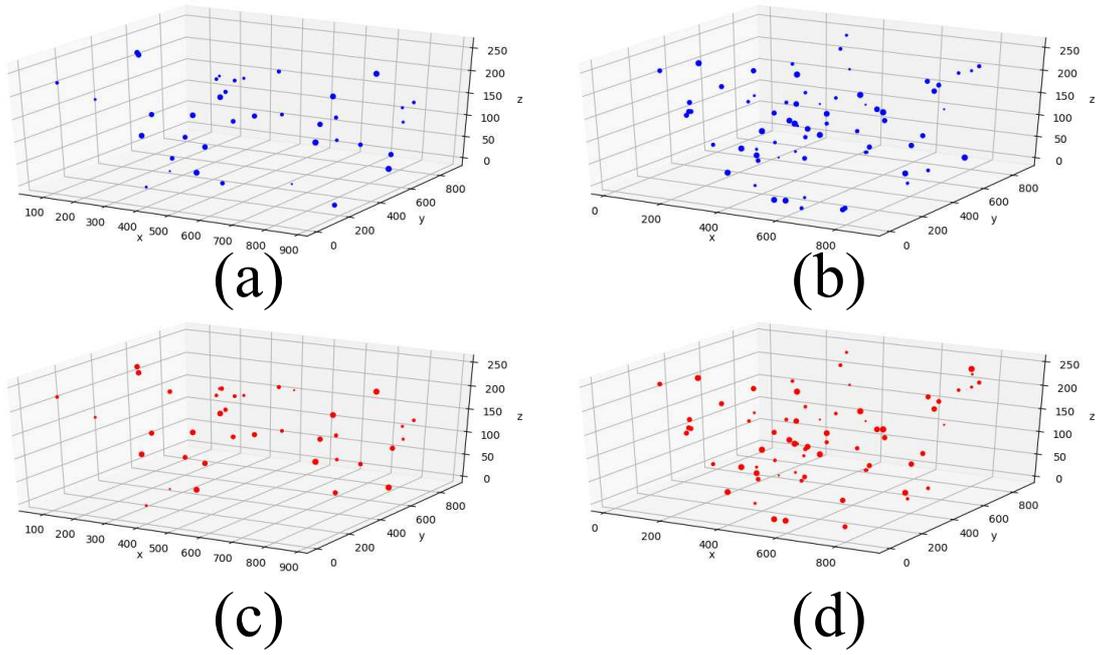}
\caption{Volume reconstructions from the predicted particle information by the DNN; (a) Original volume including 40 particles, (b) original volume including 73 particles, (c) predicted volume corresponding to (a), and predicted volume corresponding to (b).
Visualization 1 is the movie for the case of 40 particles. 
}
\label{fig:volume1}
\end{figure}

Figure \ref{fig:volume1} shows the volume reconstructions from the predicted particle information obtained by the DNN.
Figure \ref{fig:volume1}(a) shows the original volume including 40 particles and Fig. \ref{fig:volume1}(c) shows the predicted volume. 
The size of the spheres in the figures represent the particle size.
Figure \ref{fig:volume1}(b) shows the original volume including 73 particles and Fig. \ref{fig:volume1}(d) shows the predicted volume. 
The predicted volumes are generally in good agreement with the original volumes.

\begin{figure}
\centering
\includegraphics[width=\linewidth]{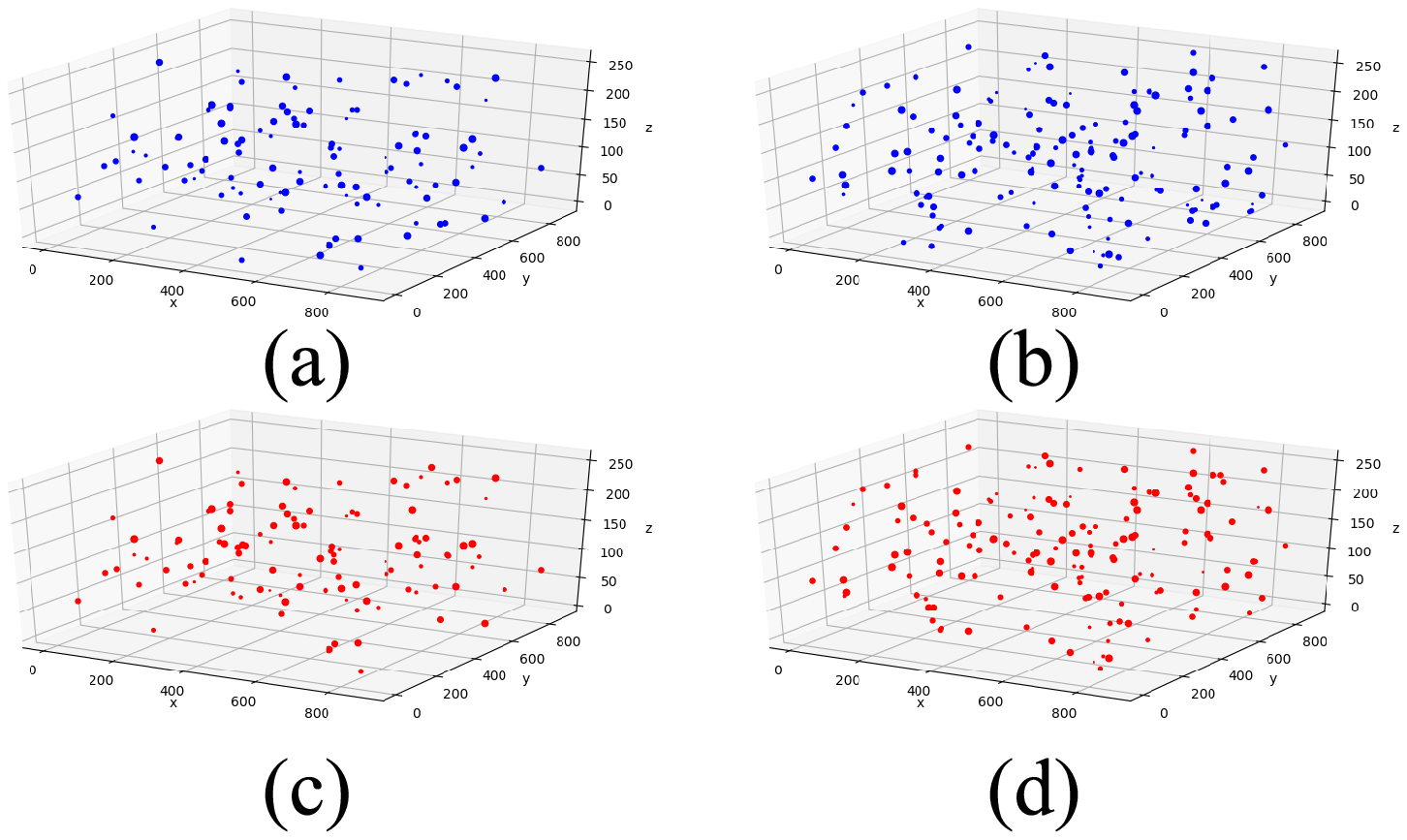}
\caption{Volume reconstructions from the predicted particle information by the DNN; (a) Original volume including 116 particles, (b) original volume including 163 particles, (c) predicted volume corresponding to (a), and predicted volume corresponding to (b).
Visualization 2 is the movie for the case of 163 particles. 
}
\label{fig:volume2}
\end{figure}

Figure \ref{fig:volume2} shows the results when the number of particles was increased.
Figure \ref{fig:volume2}(a) shows the original volume including 116 particles and Fig. \ref{fig:volume2}(c) shows the predicted volume. 
Figure \ref{fig:volume2}(b) shows the original volume including 163 particles and Fig. \ref{fig:volume2}(d) shows the predicted volume. 
The predicted volumes are again in good agreement with the original volumes.

Table \ref{tbl:error} summarizes the errors in the lateral and axial positions and the size, between the original and predicted results, when changing the number of particles. 
The errors are calculated by $1/P \sum_{j=1}^{P} |A_j-B_j|$ where $A_j$ and $B_j$ are the original and predicted values for the $j$-th particle, and $P$ is the number of particles.

In conventional digital holography, the axial resolution $\delta_z$ is subject to the numerical aperture, NA, and is calculated by $\delta_z \approx \lambda / (1.4 {\rm NA}^2)$ where $\lambda$ is the wavelength \cite{latychevskaia2016inverted}. 
In this study, to avoid aliasing error due to the sampling pitch $p$ of the camera, the spread area of diffracted light from a particle in the hologram plane is limited. The radius of the spread area is determined by $R=z \tan(\sin^{-1}(\lambda/2p))$ where $z$ is the recording distance of the particle.
Therefore, the NA is calculated by NA$=R/\sqrt{z^2+R^2}$, resulting in an axial resolution of approximately $\delta_z=0.45$ mm.
Next, we estimated the axial error of the predicted result.
The particle information has a 256-grayscale, so the axial step is $\Delta_z$=$(z_2-z_1) / 256$=(3 cm$-$1 cm) / 256 $\approx$ 0.078 mm.
For example, in the case of 40 particles, the axial error (in 256-grayscale) is 3.15. 
The error in millimeter is only $3.15 \times \Delta_z \approx 0.25$ mm.
As can be seen, the error is smaller than the theoretical axial resolution in digital holography.
The axial resolution of the reconstructed image obtained by diffraction calculation has little influence on the interference fringe of the particle. On the other hand, the DNN can detect small changes in the interference fringes of particles in the hologram, so that a high axial resolution can be achieved.

The size step is $\Delta_s=(100  \mu {\rm m} -20 \mu {\rm m})/256$ $\approx$ 0.3125 $\mu$m. 
For example, in the case of 40 particles, the size error (in 256-grayscale) is 2.82. 
The physical error in a micrometer is only $2.82 \times \Delta_s \approx 0.88$ $\mu$m. 
In addition, the lateral error (in 256-grayscale) is 0.15. 
The position error is about 0.15 $\times$ 10 $\mu$m = 1.5 $\mu$ m.
These show that the lateral position and size can also be predicted with high accuracy.

The calculation time of the prediction on the GPU is approximately 0.28 seconds per hologram.
In contrast, if we used diffraction-based particle detection, the detection requires 256 diffraction calculations.
When we used the angular spectrum method with a 1,024$\times$1,024 pixel hologram, the calculation time on the GPU is approximately 2.1 seconds. During the calculation, we used zero padding to avoid wraparound noise in the reconstructed images.
The DNN approach is effective in terms of calculation time. Visualizations 1 and 2 are movies for the cases of 40 and 163 particles, respectively.

\begin{table}[]
\centering
\caption{Errors of lateral and axial positions, and the sizes of particles.}
\label{tbl:error}
\begin{tabular}{|c|c|c|c|}
\hline
\multirow{2}{*}{Number of particles} & \multicolumn{3}{c|}{Error (in gray scale unit)} \\ \cline{2-4} 
                                     & Lateral           & Axial           & Size          \\ \hline
40                                   & 0.15              & 3.15            & 2.82          \\ \hline
73                                   & 0.096             & 3.19            & 2.31          \\ \hline
116                                  & 0.11              & 2.95            & 5.07          \\ \hline
163                                  & 0.12              & 3.41            & 5.60          \\ \hline
\end{tabular}
\end{table}

\section{Conclusions}
We proposed a DNN-based particle volume reconstruction in an in-line hologram setup. Although the in-line setup is contaminated by unwanted images that hinder the prediction,
the DNN was not affected by these  unwanted images and could predict the lateral and axial positions, and the size of particles directly from a hologram, without the need for a diffraction calculation.
The precision of the axial position predicted by the DNN exceeded the axial resolution  determined by the numerical aperture and wavelength. In addition, the calculation time of the DNN was faster than conventional diffraction-based particle detection.
The DNN approach would be very useful for particle velocimetry, the analysis, and design of fluid flow channels, and other applications.

This work was partially supported by JSPS KAKENHI Grant Numbers 16K00151.

\bibliographystyle{unsrt}
\bibliography{sample}

\begin{thebibliography}{10}

\bibitem{adams1997particle}
Mike Adams, Thomas~M Kreis, and Werner~PO J{\"u}ptner.
\newblock Particle size and position measurement with digital holography.
\newblock In {\em Optical Inspection and Micromeasurements II}, volume 3098,
  pages 234--241. International Society for Optics and Photonics, 1997.

\bibitem{murata2000potential}
Shigeru Murata and Norifumi Yasuda.
\newblock Potential of digital holography in particle measurement.
\newblock {\em Opt. Laser Technol.}, 32(7-8):567--574, 2000.

\bibitem{soulez2007inverse}
Ferr{\'e}ol Soulez, Lo{\"\i}c Denis, {\'E}ric Thi{\'e}baut, Corinne Fournier,
  and Charles Goepfert.
\newblock Inverse problem approach in particle digital holography: out-of-field
  particle detection made possible.
\newblock {\em JOSA A}, 24(12):3708--3716, 2007.

\bibitem{pan2003digital}
Gang Pan and Hui Meng.
\newblock Digital holography of particle fields: reconstruction by use of
  complex amplitude.
\newblock {\em Appl. Opt.}, 42(5):827--833, 2003.

\bibitem{ohman2016off}
Johan {\"O}hman and Mikael Sj{\"o}dahl.
\newblock Off-axis digital holographic particle positioning based on
  polarization-sensitive wavefront curvature estimation.
\newblock {\em Appl. Opt.}, 55(27):7503--7510, 2016.

\bibitem{latychevskaia2014holographic}
Tatiana Latychevskaia and Hans-Werner Fink.
\newblock Holographic time-resolved particle tracking by means of
  three-dimensional volumetric deconvolution.
\newblock {\em Opt. Express}, 22(17):20994--21003, 2014.

\bibitem{tanaka2016phase}
Yohsuke Tanaka, Shunsuke Tani, and Shigeru Murata.
\newblock Phase retrieval method for digital holography with two cameras in
  particle measurement.
\newblock {\em Opt. Express}, 24(22):25233--25241, 2016.

\bibitem{yevick2014machine}
Aaron Yevick, Mark Hannel, and David~G Grier.
\newblock Machine-learning approach to holographic particle characterization.
\newblock {\em Opt. Express}, 22(22):26884--26890, 2014.

\bibitem{schneider2016fast}
Bendix Schneider, Joni Dambre, and Peter Bienstman.
\newblock Fast particle characterization using digital holography and neural
  networks.
\newblock {\em Appl. Opt.}, 55(1):133--139, 2016.

\bibitem{ronneberger2015u}
Olaf Ronneberger, Philipp Fischer, and Thomas Brox.
\newblock U-net: Convolutional networks for biomedical image segmentation.
\newblock In {\em International Conference on Medical image computing and
  computer-assisted intervention}, pages 234--241. Springer, 2015.

\bibitem{goodfellow2014generative}
Ian Goodfellow, Jean Pouget-Abadie, Mehdi Mirza, Bing Xu, David Warde-Farley,
  Sherjil Ozair, Aaron Courville, and Yoshua Bengio.
\newblock Generative adversarial nets.
\newblock In {\em Advances in neural information processing systems}, pages
  2672--2680, 2014.

\bibitem{dardikman2018phase}
Gili Dardikman and Natan~T Shaked.
\newblock Phase unwrapping using residual neural networks.
\newblock In {\em Computational Optical Sensing and Imaging}, pages CW3B--5.
  Optical Society of America, 2018.

\bibitem{Kingma2014Adam}
Diederik~P Kingma and Jimmy Ba.
\newblock Adam: A method for stochastic optimization.
\newblock {\em arXiv preprint arXiv:1412.6980}, 2014.

\bibitem{shimobaba2012computational}
Tomoyoshi Shimobaba, Jiantong Weng, Takahiro Sakurai, Naohisa Okada, Takashi
  Nishitsuji, Naoki Takada, Atsushi Shiraki, Nobuyuki Masuda, and Tomoyoshi
  Ito.
\newblock Computational wave optics library for {C++}: {CWO++} library.
\newblock {\em Comput. Phys. Commun.}, 183(5):1124--1138, 2012.

\bibitem{chollet2015keras}
Fran\c{c}ois Chollet et~al.
\newblock Keras.
\newblock \url{https://keras.io}, 2015.

\bibitem{latychevskaia2016inverted}
Tatiana Latychevskaia and Hans-Werner Fink.
\newblock Inverted gabor holography principle for tailoring arbitrary shaped
  three-dimensional beams.
\newblock {\em Sci. Rep.}, 6:26312, 2016.

\end{thebibliography}

\end{document}